\documentclass[iop]{emulateapj-rtx4}
\usepackage{epsfig}
\usepackage{graphicx}
\usepackage{longtable}
\usepackage{natbib}
\usepackage{subfigure}
\usepackage{xcolor}
\usepackage{amsmath}
\usepackage{multirow}

\newcommand{\Msolar}{M$_{\odot}$}

\newcommand{\Rsolar}{R$_{\odot}$}
\newcommand{\kms}{km s$^{-1}$}

\newcommand{\numax}{$\nu_{\mathrm{max}}$}
\newcommand{\dnu}{$\Delta\nu$}

\shorttitle{An Evolved Blue Straggler in M67}
\shortauthors{Leiner et al.}

\begin{document}

\title{The K2 M67 Study: An Evolved Blue Straggler in M67 from K2 Mission Asteroseismology}

\author{Emily Leiner\altaffilmark{1}, Robert
  D. Mathieu\altaffilmark{1}, Dennis Stello \altaffilmark{2,3,4}, and Andrew Vanderburg\altaffilmark{5}}
\email{leiner@astro.wisc.edu}
St

\altaffiltext{1}{Department of Astronomy, University of Wisconsin-Madison, 475 North Charter St, Madison, WI 53706, USA}
\altaffiltext{2}{Sydney Institute for Astronomy (SIfA), School of Physics, University
  of Sydney, Sydney, NSW 2006, Australia}
\altaffiltext{3}{Stellar Astrophysics Centre, Department of Physics and Astronomy, Aarhus University, Ny Munkegade 120, DK-8000 Aarhus C, Denmark}
\altaffiltext{4}{School of Physics, University of New South Wales, NSW 2052, Australia}
\altaffiltext{5}{Harvard--Smithsonian Center for Astrophysics, 60 Garden St., Cambridge, MA 02138, USA}

\begin{abstract}
Yellow straggler stars (YSSs) fall above the
subgiant branch in optical color-magnitude diagrams, between the blue
stragglers and the red giants. YSSs may represent a population of evolved blue stragglers, but none have the direct and precise mass and radius measurements needed to determine their evolutionary states and formation histories. Here we report the first asteroseismic mass and radius measurements of such a star, the yellow straggler S1237 in the open cluster M67. We apply asteroseismic scaling relations to a frequency analysis of the \textit{Kepler} K2 light curve and find a mass of $2.9 \pm 0.2$~\Msolar~and a radius of $9.2 \pm 0.2$ ~\Rsolar. This is more than twice the mass of the main-sequence turnoff in M67, suggesting S1237 is indeed an evolved blue straggler. S1237 is the primary in a spectroscopic binary. We update the binary orbital solution and use  spectral energy distribution (SED) fitting to constrain the color-magnitude diagram (CMD) location of the secondary star. We find that the secondary is likely an upper main-sequence star near the turnoff, but a slightly hotter blue straggler companion is also possible. We then compare the asteroseismic mass of the primary to its mass from CMD fitting, finding the photometry implies a mass and radius more than $2\sigma$ below the asteroseismic measurement. Finally, we consider formation mechanisms for this star and suggest that S1237 may have formed from dynamical encounters resulting in stellar collisions or a binary merger.
\end{abstract}

\section{Introduction}
In color-magnitude diagrams (CMDs) of open clusters, stars are observed between the blue straggler stars (BSSs) and the
red giant branch (RGB). We call these yellow straggler stars (YSSs). In M67, a $\sim$4 Gyr, solar-metallicity open cluster \citep{Taylor2007, Montgomery1993}, there are four 3D kinematic members found in this CMD region. Three are binary systems, and one is a single star \citep{Geller2015}. While some YSSs might be
explained as the combined light of two cluster stars (e.g., a red giant-blue
straggler binary), many may be evolved BSSs  \citep{Mathieu1990}. These stars would be post-main-sequence stars more
massive than the main-sequence turnoff (MSTO). 

BSSs are thought to form from mass transfer in binary systems \citep{McCrea1964, Gosnell2014}, stellar collisions during dynamical encounters \citep{Leonard1989}, or binary mergers induced by Kozai cycles \citep{Perets2009}. In M67, one of the four YSSs has a helium white dwarf (WD) companion, suggesting it is an evolved BSS formed from mass transfer \citep{Landsman1997}. 

Asteroseismic analysis of \textit{Kepler} stars have uncovered
red giants in open clusters more massive than the
cluster standard \citep{Stello2011,Brogaard2012, Corsaro2012}. These stars are called evolved blue straggler stars (E-BSSs). They fall in or near their clusters' red clumps, but YSSs may be bluer examples of the E-BSSs, making them more identifiable in cluster CMDs. Mass measurements for YSSs are needed to confirm this hypothesis. 

Recent data from the \textit{Kepler} K2 mission allow for the asteroseismic measurement of stellar masses and radii in M67 \citep{Stello2016b}, providing an exciting opportunity to study the YSSs in this cluster. Here we report the first determination of an asteroseismic mass and radius for the primary star in the M67 binary YSS S1237 \citep{Sanders1977}. We summarize radial-velocity (RV) and X-ray observations of S1237, and use the 
spectral energy distribution (SED) to constrain the mass of the
secondary star. We compare the asteroseismic mass to the primary mass implied from its CMD position, and comment on possible formation pathways for this star. 

\section{Asteroseismic Measurements}\label{Section:Asteroseimology}

M67 was observed during Campaign 5 of the \textit{Kepler} K2 mission, providing high-precision 75-day light curves for targets in the cluster. Light curves were obtained using the method of \citet{Vanderburg2014}. 

The light curve of YSS S1237 shows solar-like oscillations and our seismic analysis followed the same approach as \citet{Stello2016b}. We processed the light curve of S1237 as in \citet{Stello2015} and used
the pipeline of \citet{Huber2009} to extract the large frequency separation ($\Delta\nu$)) and frequency of maximum power
(\numax)~from the Fourier frequency spectra of the light curve. From these global frequencies, we used the asteroseismic scaling
relations, \dnu\ $\propto \sqrt{M/R^3}$ and \numax\ $\propto
M/(R^2\sqrt{T_{\mathrm{eff}}}$) (e.g. \citealt{Kjeldsen1995}), to determine a mass and radius for the
star. \

\begin{figure}
\includegraphics[width=.95\linewidth, angle=0]{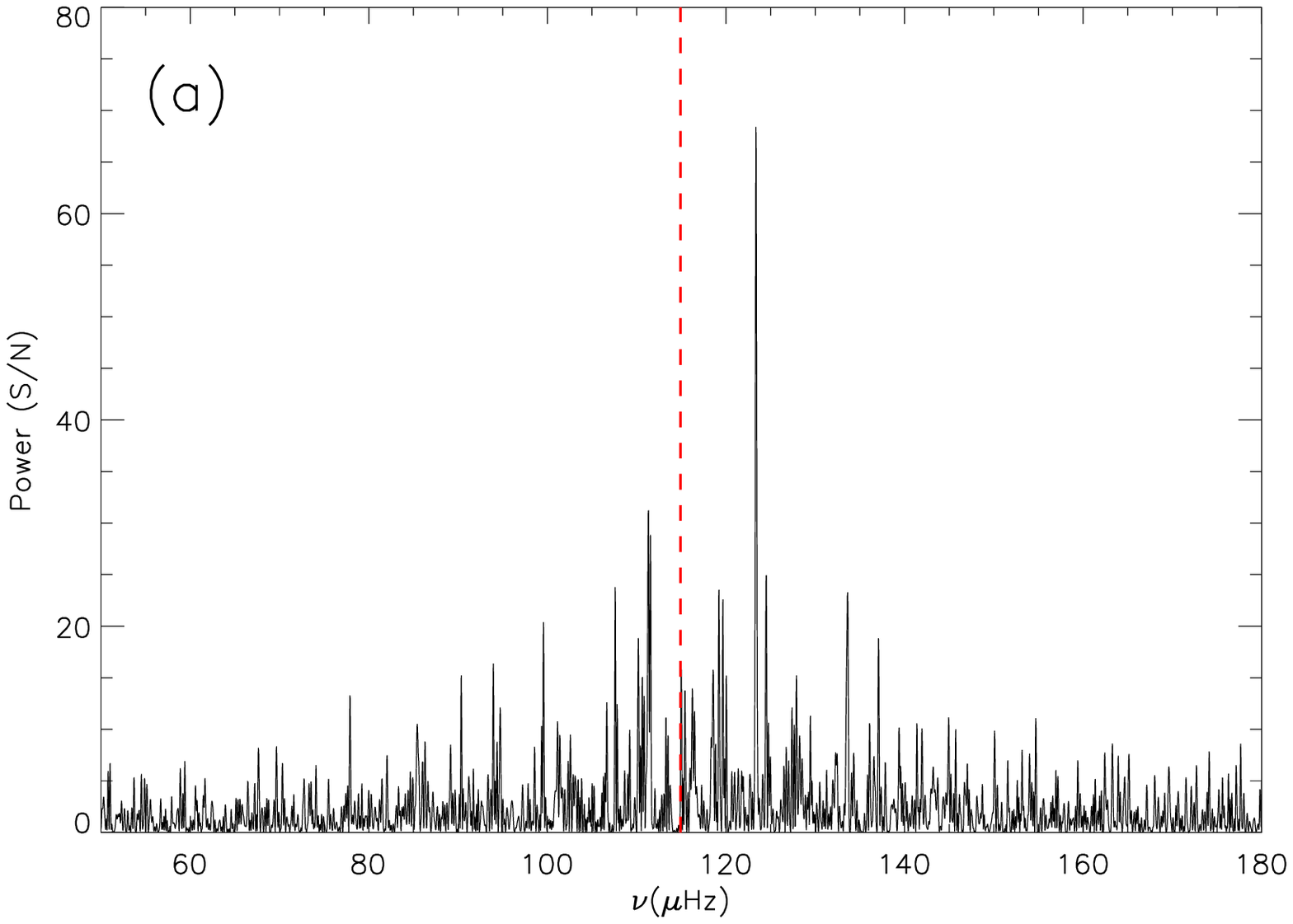}
\includegraphics[width=.95\linewidth, angle=0]{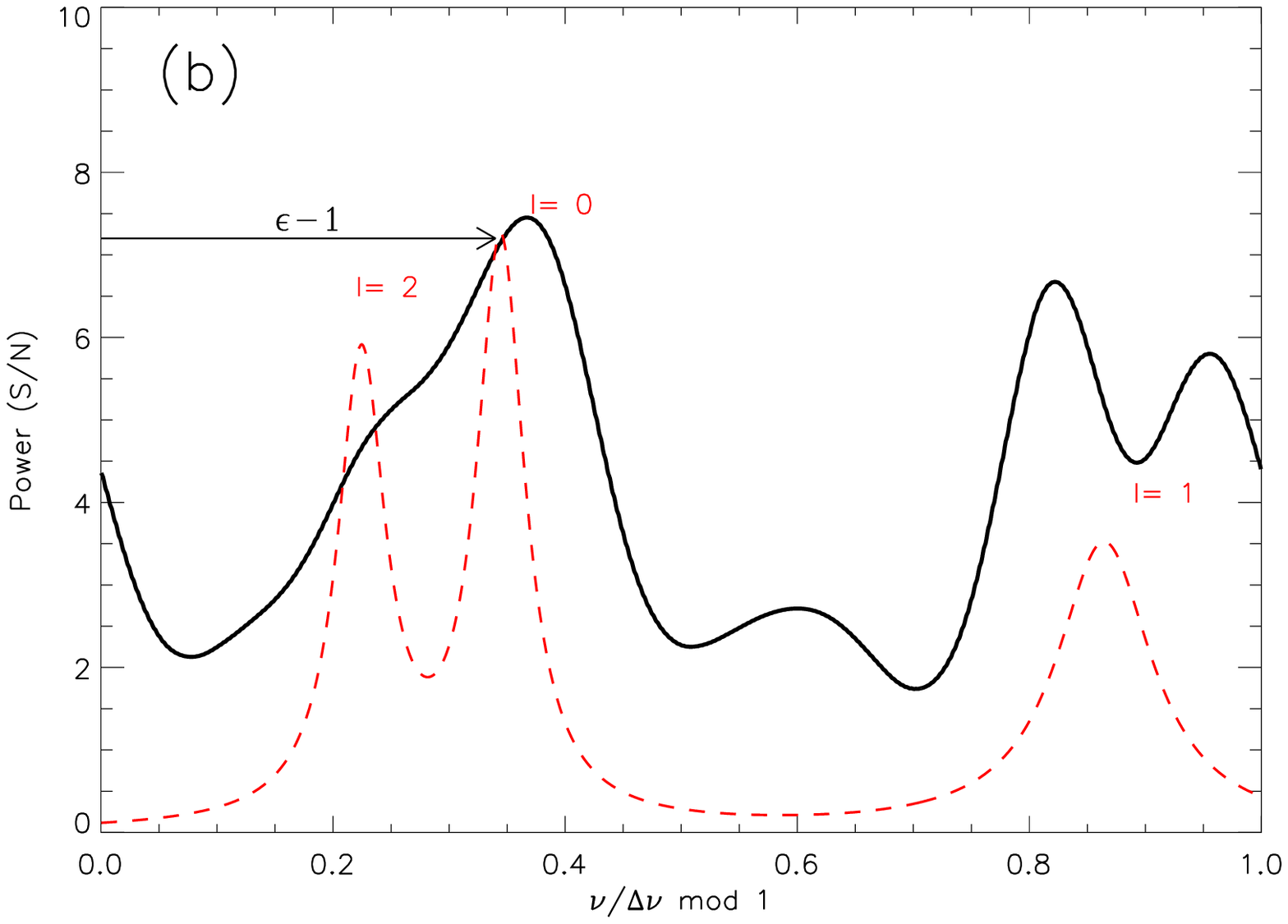}
\caption{\textbf{(a)} Background corrected power spectrum of
 S1237. The dashed red line shows the location of $\nu_{\text{max}}$. \textbf{(b)} Folded and smoothed spectrum of the central $\pm 2
 \Delta\nu$ range around $\nu_{\text{max}}$ (black curve).  The red dashed line shows an empirical model representing the average of a large ensemble of red giants used to measure $\epsilon$ and determine the location of the different modes \citep{Stello2016,Stello2016b}. The arrow indicates the offset from zero of the radial modes. \label{PS}}
\end{figure}

%\subsection{Scaling Relations}

\begin{table}
\begin{center}
\caption{Parameters for S1237}\label{asteroparams}
\begin{tabular}{lll}
\hline
\hline
\multicolumn{3}{c}{General}\\
\hline
RA & 8:51:50.20&\\
Dec. & 11:46:07.0&\\
EPIC ID & 211408357 &\\
WOCS ID\tablenotemark{1} & 1015 &\\
Sanders ID\tablenotemark{2} & S1237&\\
\hline
\hline
\multicolumn{3}{c}{Updated Binary Orbital Parameters}\\
\hline
Period (P) (days) & $698.4$ & $\pm 0.3$\\
Eccentricity (e) & 0.087 & $\pm 0.015$\\
Amplitude (K) (\kms) & 5.024 & $\pm 0.071$ \\
f(m) (\Msolar)& 0.0091 & $\pm 0.0004 $\\
\hline
\hline
\multicolumn{3}{c}{Old Binary Orbital Paramters\tablenotemark{3}}\\
\hline
Period (P) (days) & $697.8$ & $\pm 0.7$\\
Eccentricity (e) & 0.105 & $\pm 0.015$\\
Amplitude (K) (\kms) & 5.03 & $\pm 0.07$ \\
f(M)) (\Msolar)& 0.0091 & $\pm 0.0004 $\\
\hline
\hline
\multicolumn{3}{c}{Measured $\Delta\nu$-scaling}\\
\hline
$\nu_{\text{max}}$ ($\mu$Hz) & 114.89 & $\pm 2.06$\\
$\Delta\nu$ ($\mu$Hz) & 8.32& $\pm 0.11$\\
Radius (\Rsolar) & 9.11 & $\pm 0.18$ \\
Mass (\Msolar) & 2.87 & $\pm 0.23$\\
Surface gravity (log $g$) & 2.98 & $\pm$ 0.01 \\
\hline
\hline
\multicolumn{3}{ c}{Corrected $\Delta\nu$-scaling assuming RGB}\\ \hline
$f_{\Delta\nu}$ & 1.0057 & \\
Radius (\Rsolar) & 9.21& $\pm$ 0.19\\
Mass (\Msolar) & 2.94& $\pm$ 0.24\\
\hline
\hline
\multicolumn{3}{ c}{Corrected $\Delta\nu$-scaling assuming HeB} \\ \hline
$f_{\Delta\nu}$ & 1.0089 & \\
Radius (\Rsolar) & 9.27& $\pm$ 0.19\\
Mass (\Msolar) & 2.97 & $\pm$ 0.24\\
\hline
%\hline
%&Photometry&\\
%\hline 
%
%%FUV & 22.268 & $ \pm 0.473$ \\
%NUV & 17.673& $ \pm .032$\\
%B & 11.72 & \nodata \\
%V & 10.78 & \nodata\\
%I & 9.84 & \nodata \\
%J & 9.124 & $\pm 0.027 $\\
%H & 8.661 & $\pm 0.026 \\
%K & 8.554 & $\pm$ 0.018\\
%W1 & 8.477 &$\pm$ 0.021\\
%W2 & 8.571 & $\pm 0.02$\\ 
%W3 & 8.501 & $\pm 0.022$\\
%\hline
\end{tabular}
\end{center}
\tablenotetext{1}{from \citet{Geller2015}}
\tablenotetext{2}{from \citet{Sanders1977}}
\tablenotetext{3}{from \citet{Mathieu1990}}

\end{table}

\begin{equation}
\frac{M}{M_{\odot}}= \bigg(\frac{\Delta\nu}{\Delta\nu_{\odot}}\bigg)^{-4}\bigg(\frac{\nu_{\text{max}}}{\nu_{\text{max}_{\odot}}}\bigg)^{-3}\bigg(\frac{T_{\text{eff}}}{T_{\text{eff}_{\odot}}}\bigg)^{\frac{3}{2}}
\end{equation}

\begin{equation}
\frac{R}{R_{\odot}}= \bigg(\frac{\Delta\nu}{\Delta\nu_{\odot}}\bigg)^{-2}\bigg(\frac{\nu_{\text{max}}}{\nu_{\text{max}_{\odot}}}\bigg)\bigg(\frac{T_{\text{eff}}}{T_{\text{eff}_{\odot}}}\bigg)^{\frac{1}{2}}
\end{equation} 

The \dnu\ scaling relation needs a small metallicity- and
temperature-dependent correction (e.g.~\citealt{White2011}). To derive this
correction factor, $f_{\Delta\nu}$, we adopt $T_{\text{eff}}= 4997 \pm 91$ K and [Fe/H]= 0.072 $\pm$ 
0.033 (from APOGEE\footnote{http://www.sdss.org/surveys/apogee/};
\citealt{APOGEE_DR12}), which we feed into the correction
interpolator of \citet{Sharma2016}. %[ApJ,822,15] 
We derive two correction factors, one assuming the star is helium-core
burning (HeB)
and the other assuming it is an RGB
star. We obtain R = 9.27 $\pm$ 0.19~\Rsolar~and M = 2.97 $\pm$ 0.24 ~\Msolar~for a
HeB star, or R = 9.21 $\pm$ 0.19~\Rsolar~and
M = 2.94 $\pm$ 0.24~\Msolar~for an RGB star (Table~\ref{asteroparams}). The errors on mass and radius come from propagating errors on $T_{\text{eff}}$, \dnu\ ,and \numax\ through Equations 1 and 2.

Our results seem to indicate this is a HeB star. A 2.6-2.9 \Msolar~star with this radius would be in the Hertzsprung Gap, and thus is unlikely
to be observed. The power 
spectrum also shows broader peaks and more blended $l= 0$ and $l= 2$
modes than typical of RGB stars. We show the full power spectrum in 
Figure~\ref{PS}a, as well as the 
central $\pm 2 \Delta\nu$ around $\nu_{\text{max}}$ folded on $\Delta\nu$ and
smoothed with a Gaussian as in \cite{Stello2016} (Figure~\ref{PS}b). The folded spectrum shows 
a broad series of multiple peaks around the location of the dipole ($l=1$)
mode. This is indicative of a large period spacing between 
dipole mixed modes \citep{Bedding2011} and of a larger coupling between the
envelope and core of the star \citep{Dupret2009}, both suggestive of a HeB
star. 
In Figure~\ref{PS}b we also measure $\epsilon$, the offset from zero of the radial
modes ($l=0$), by correlating the spectrum with a model profile following \citet{Stello2016}.
We find $\epsilon = 1.34$, outside the range for an RGB star (\citealt{Stello2016}, Fig. 2).

\section{Other Observations}
\subsection{Updated Orbit}

%\begin{figure}
%\begin{center}
%\includegraphics[width=.9\linewidth, angle=0]{1015.sb1.cut.eps}
%\subfigure{\includegraphics[width=0.7\linewidth, angle=0]{colorbar.eps}}
%\caption{Orbital solution for S1237 (WOCS 1015) \label{orbit}}
%\end{center}
%\end{figure}d

Updating the orbital solution from \citet{Mathieu1990} with additional RVs from the WIYN Open Cluster Study (WOCS; \citealt{Mathieu2000}), we find the orbital parameters do not change significantly (see Table 1). The system is a near-circular 698-day period spectroscopic binary. 

The system is single-lined in the WOCS spectra ($R \sim15,000$). Given the small amplitude of the orbit and the fact that any MS companion would be more than 2 magnitudes fainter than a $\sim9.0$ \Rsolar~giant primary, we do not expect to see the features of the secondary at the signal to noise of the WOCS spectra,. Higher resolution, high signal-to-noise spectra may allow detection of the secondary. For a primary of 2.9 \Msolar, the binary mass function ($f(m_\text{1}, m_\text{2}) =0.0091$) indicates a secondary with $m_\text{2} \geq 0.46$ \Msolar. 

\subsection{X-ray Observations}

S1237 has a 0.3-7 KeV luminosity of L$_x$= 5.5 $\times 10^{29}$ ergs s$^{-1}$, comparable to what is expected from rapid rotation \citep{vandenBerg2004}. However, WOCS spectra of the primary provide an upper limit on the $v$ sin $i$ of 10 \kms~set by the instrumental broadening and the binary is too wide for rapid rotation to be expected from tidal synchronization. \citet{Belloni1998} also detect S1237 as an X-ray source and find no satisfactory explanation. We discuss a possible explanation for this X-ray emission in Section 6.

\subsection{SED Fitting}
\begin{figure*}
\begin{center}
\subfigure[]{\includegraphics[width=.48\linewidth, angle=0]{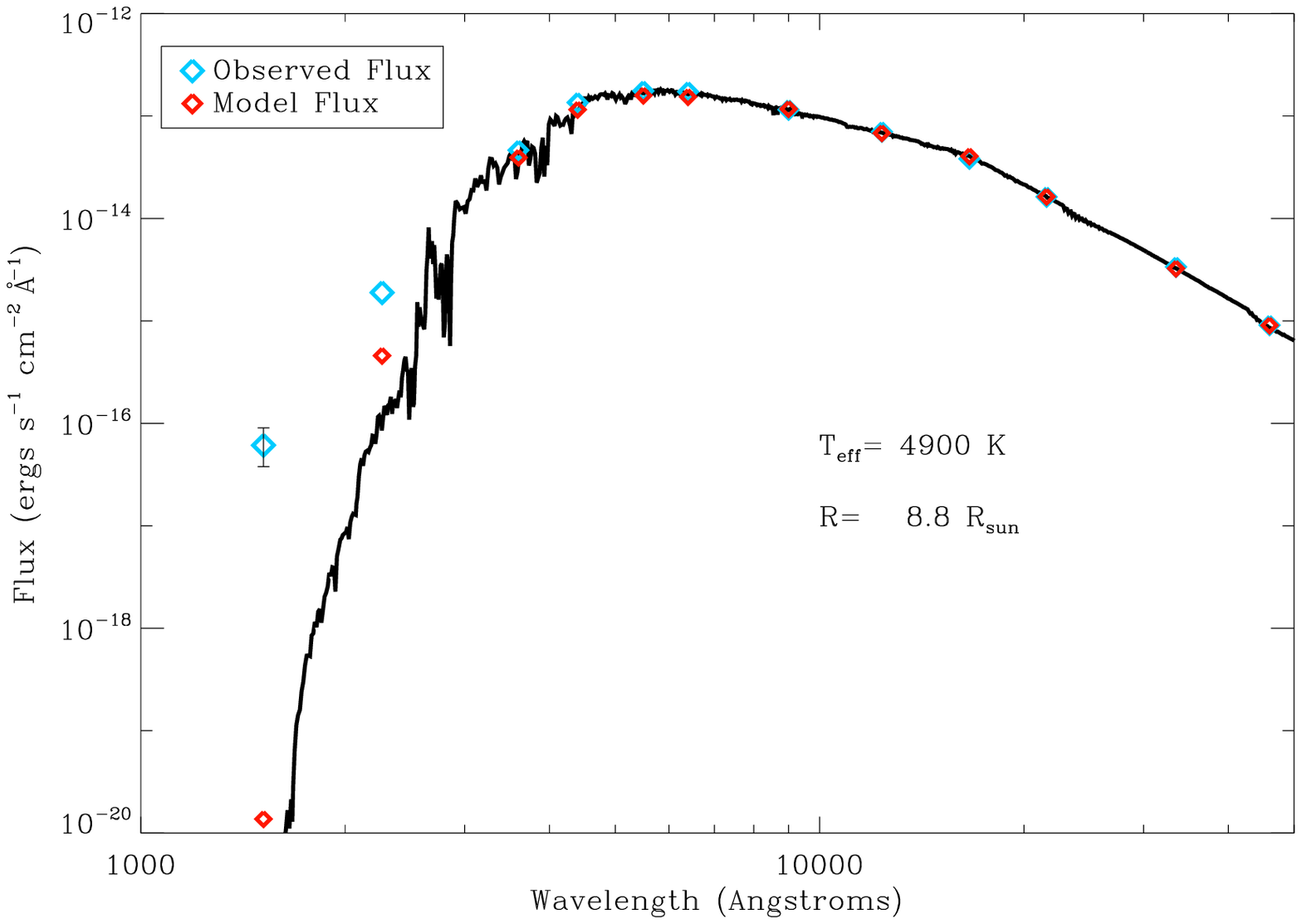}}
\subfigure[]{\includegraphics[width=.48\linewidth, angle=0]{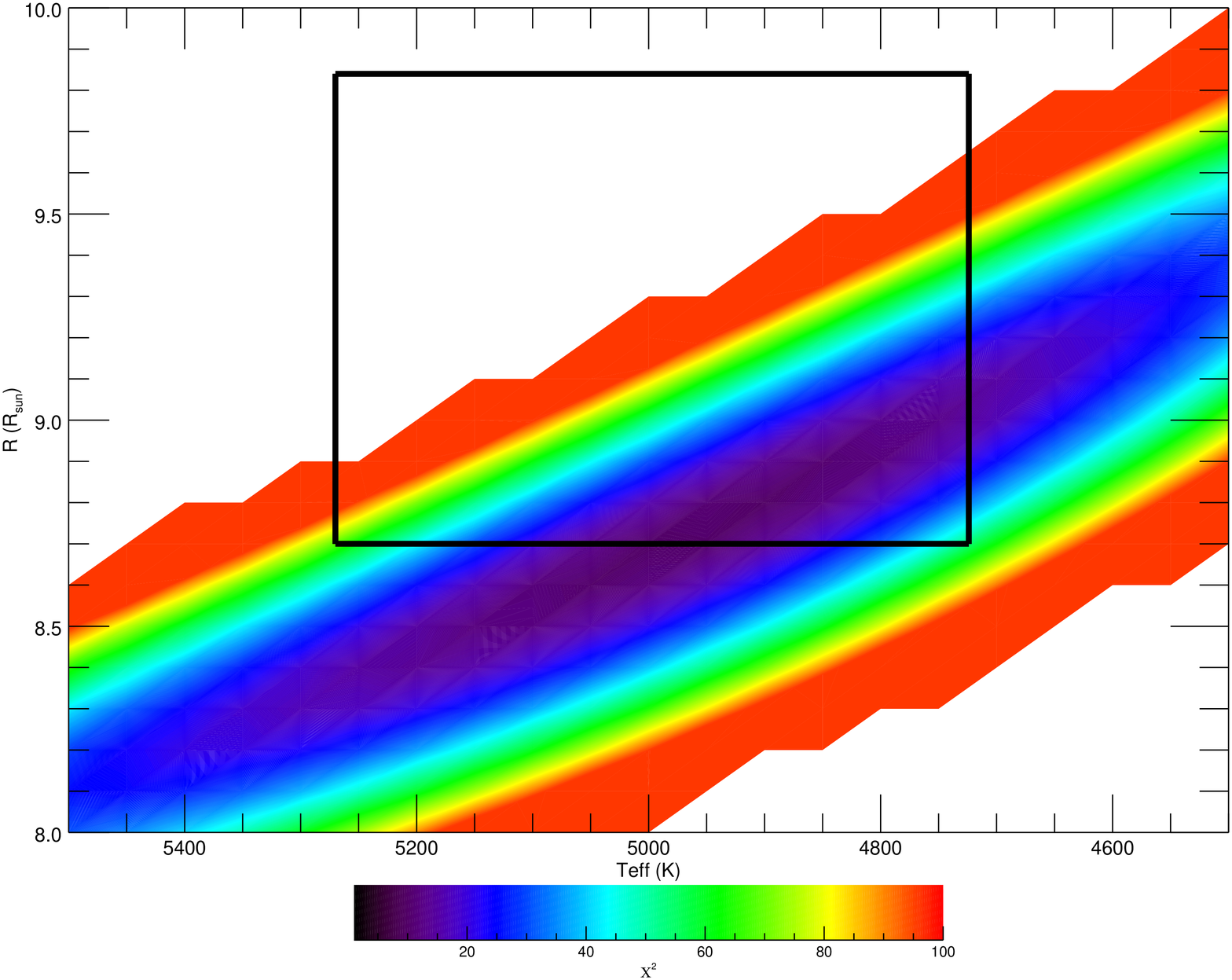}}
\subfigure[]{\includegraphics[width=.48\linewidth, angle=0]{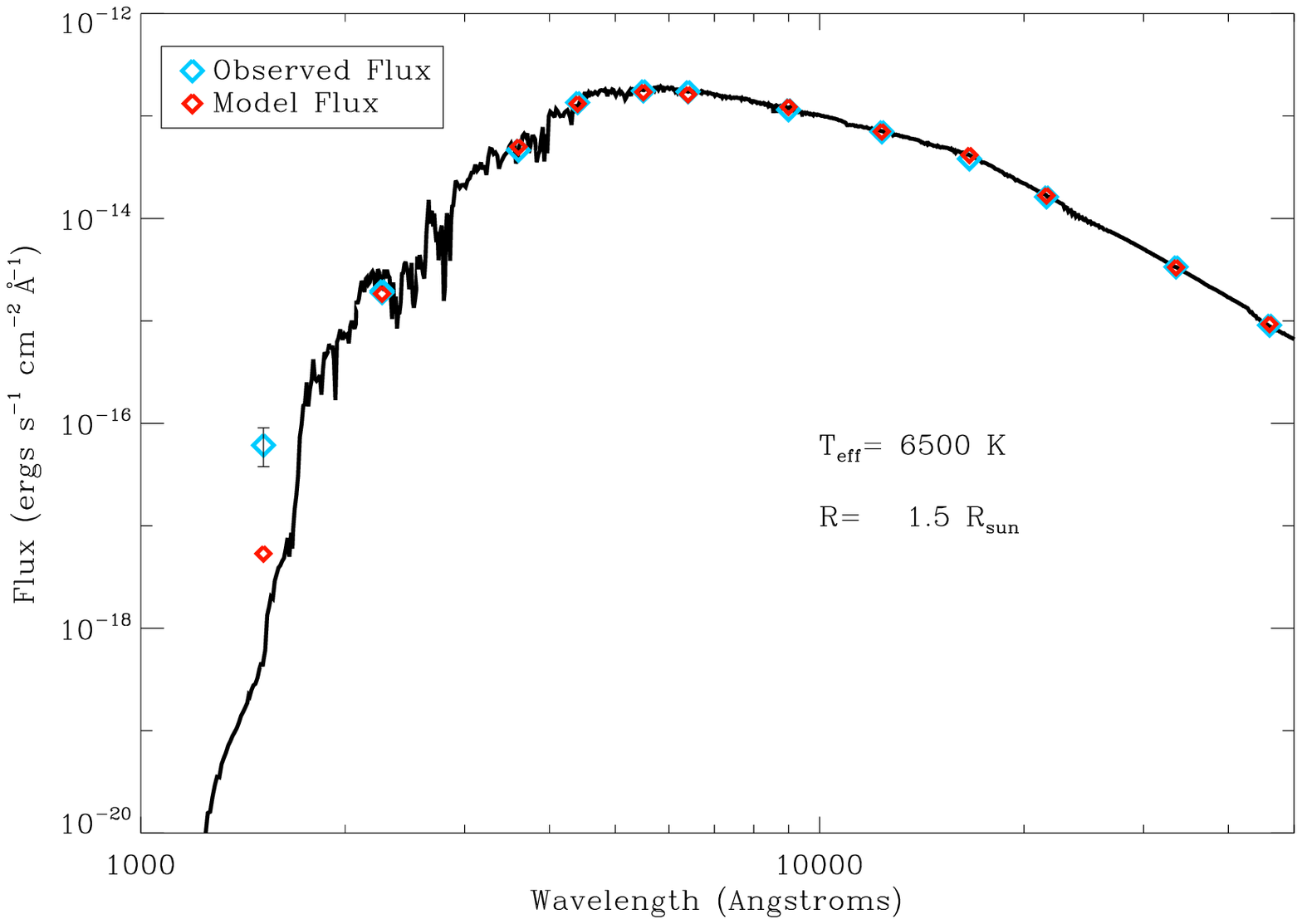}}
\subfigure[]{\includegraphics[width=.48\linewidth, angle=0]{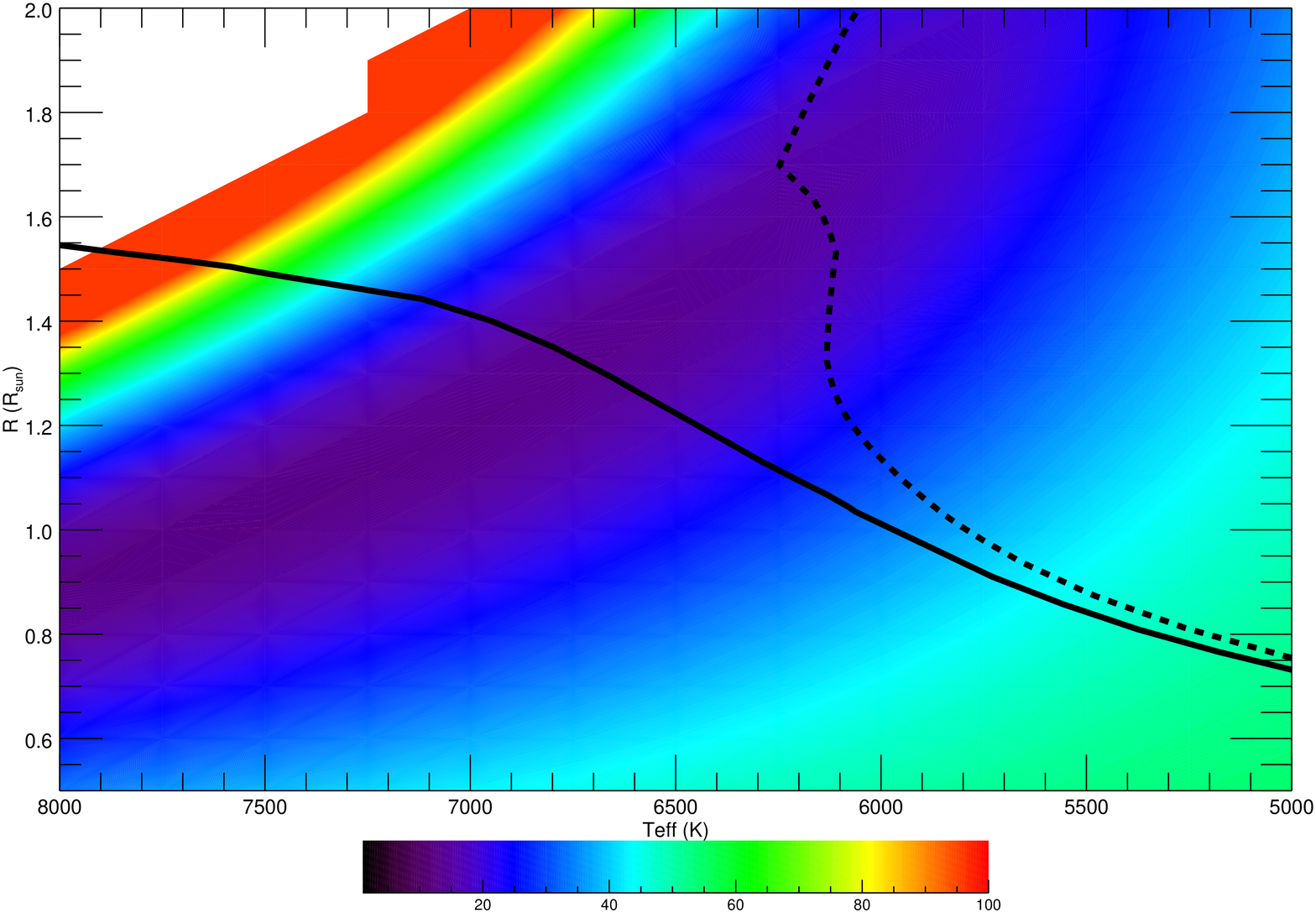}}

\caption{\textbf{(a)} The observed SED of S1237 (blue points) compared to our best fitting model of a 8.8 \Rsolar, 4900 K giant (black line). The red points indicate the model flux when the SED is convolved with the filter transmission functions. \textbf{(b)} A $\chi^2$ map resulting from fitting a grid of models for the primary to the 2MASS and WISE observations of S1237. The color indicates the $\chi^2$ value from smallest (purple) to largest (red). The black box bounds the area within $3\sigma$ of the APOGEE $T_{\text{eff}}$ and the asteroseismic radius. \textbf{(c)} The observed SED of S1237 compared to a model of the combined light of a 4900 K, 8.8 \Rsolar~ giant and a 6500 K, 1.5 \Rsolar~ MS star. \textbf{(d)} A $\chi^2$ map resulting from fitting a grid of models containing a 4900 K, 8.8 \Rsolar~ giant and various MS secondaries.. The solid black line indicates the zero-age main-sequence, and the dashed black line is a 4 Gyr isochrone.}
\end{center}
\end{figure*}

We have assembled an SED from existing UBVI photometry \citep{Montgomery1993}, the Two-Micron All Sky Survey (2MASS; \citealt{2MASS2006}), Wide Field Infrared Explorer (WISE; \citealt{Wright2010}), and the Galaxy Evolution Explorer (GALEX; \citealt{Martin2005}). 

We use $\chi^2$ minimization to fit the observed SED to a grid of models. We first fit a model of the primary to the 2MASS and WISE data. These IR measurements should be dominated by the flux from the primary, whereas the optical and UV flux may be contaminated by flux from a MS secondary. We model the primary with a 4500-5500 K giant spectrum \citep{CastelliKurucz} in steps of 50 K in order to bracket the the APOGEE $T_{\text{eff}}$ of $4997 \pm 91$ K. We use a grid of radii between 8.0-10.0 \Rsolar~in 0.1 \Rsolar~increments. We fix the surface gravity to log g= 3.0 from asteroseismic scaling (Table 1). We use distances of either 800, 850, or 900 pc based on the range given in the literature \citep{Geller2015}.  We adopt E(B-V)= 0.041 \citep{Taylor2007} and the extinction curve of \citet{Cardelli1989}. 

The best-fit SED temperature and radius only fall within the 3$\sigma$ errors on the APOGEE temperature and the asteroseismic radius if we use a distance of 900 pc. Using 900 pc, we find a best-fit of 8.8 \Rsolar and 4900 K (see Figure 2b). Using this model for our primary, we find a significant flux excess over the model in all of the bluer bands (FUV, NUV, U, B, V) indicating the presence of a secondary (see Figure 2a). Fixing the primary to these best-fit values, we model the secondary using a fixed surface gravity of log g= 4.5 (typical for a MS star), a grid of temperatures between 5000-8000 K in steps of 250 K, and secondary radii between 0.5 \Rsolar-2.0 \Rsolar~in steps of 0.1 \Rsolar. 

We find that a variety of secondary models result in $\chi^2$ values near the minimum with temperatures of 6250-8000 K and radii of 0.5-1.8 \Rsolar, but if we restrict the radius to fall between the ZAMS radius and the radius at 4 Gyr for a given temperature, we find the best fit to be a $\sim$6250-6750 K, 1.2-1.6 \Rsolar~star (Figure 2c, d). This range includes both stars on the upper main-sequence in M67, as well as blue straggler stars near the MSTO. For simplicity, we assume the star is on the upper MS when considering formation scenarios, but the possibility that the secondary is a blue straggler is worth follow up, i.e. with higher resolution spectra that could detect the secondary. 

We note there is a GALEX FUV excess in the SED (Figure 2c), but we do not consider this excess to be meaningful. Castelli-Kurucz models have poor resolution in the FUV, and a few unresolved emission lines can change the flux substantially. In fact, MS stars near the turn off in M67 have widely varying GALEX FUV fluxes, and several are observed to have comparable FUV magnitudes.

\section{Photometric versus Asteroseismic Mass}

\begin{figure*}[htbp]
\begin{center}
\includegraphics[width=.95\linewidth, angle=0]{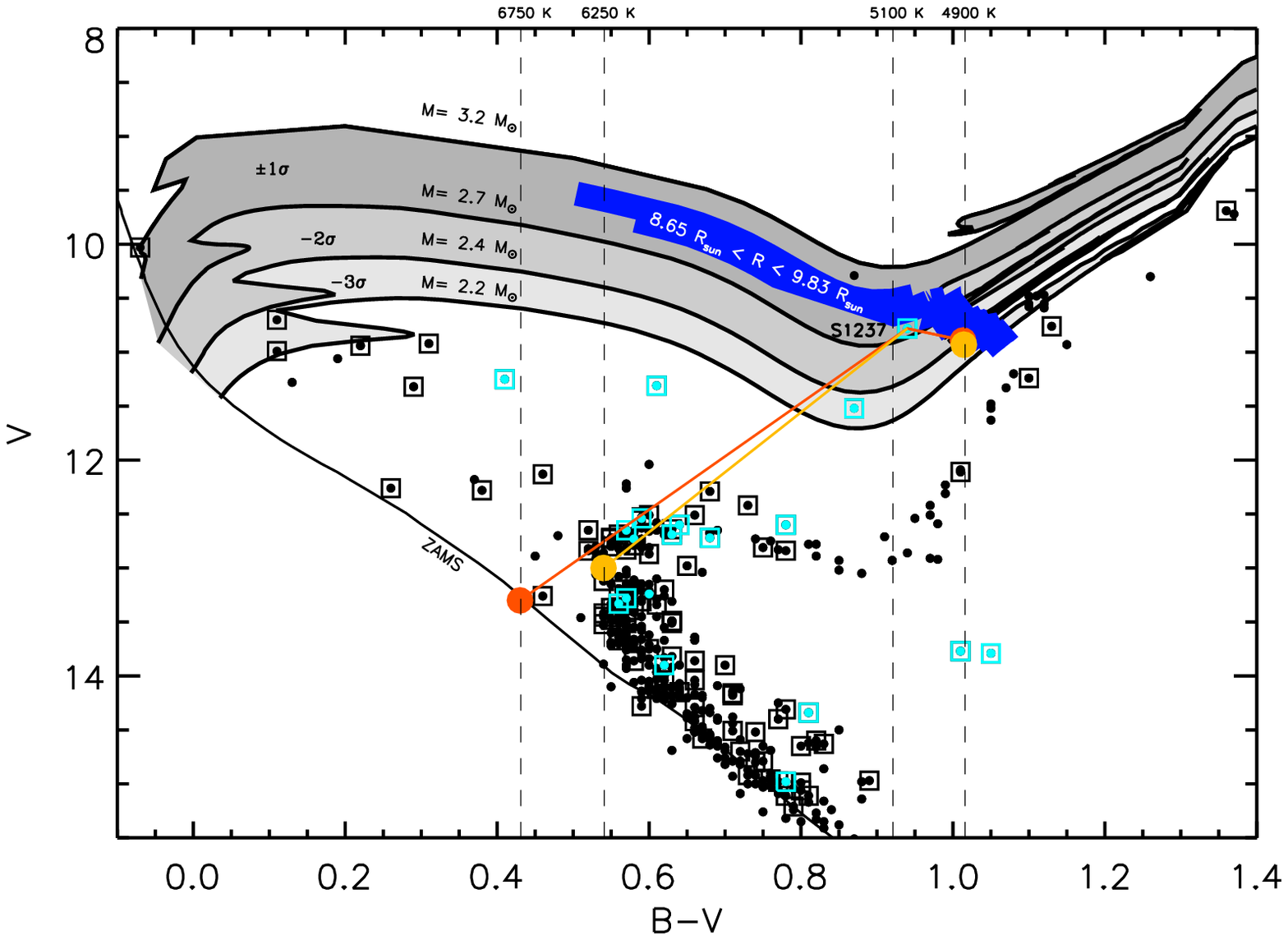}
\caption{A CMD of M67. We plot X-ray sources from
 \citet{Belloni1998} and \citet{vandenBerg2002} in light blue and all other 3D
 kinematic members in black. Binaries are boxed \citep{Geller2015}. We
 use d= 900 pc and E(B-V)= 0.041. Also shown are MESA evolutionary tracks for a 3.2, 2.7, 2.4,
 and 2.2 \Msolar. The shaded regions between
 these tracks indicate the offset from the
 asteroseismic mass: the darkest gray is within $1\sigma$, medium gray within $2\sigma$, and lightest gray within $3\sigma$. The blue 
 region indicates the points on the evolutionary tracks where the star has a
 radius within $3\sigma$ of the asteroseismic values. We show two examples of
 the possible photometric deconvolution, one with a BSS secondary (red
 circles) and one with a MS secondary near the turnoff (orange
 circles). The dashed vertical lines show the $1\sigma$ range in temperature from
 APOGEE spectra for the primary and SED fitting for the
 secondary \citep{Torres2010}. \label{CMD}}
\end{center}
\end{figure*}

In Figure~\ref{CMD} we decompose the light of the binary to show the location of the primary and the secondary in a CMD. We show two possible deconvolutions: an MS secondary and a BSS secondary. The secondary is fixed to either a 6750 K BSS (red circle), or a 6250 K MS star near the turnoff (orange circle). The primary is a $\sim4900$ K giant. 

We compare the CMD position of the primary and secondary to evolutionary tracks using Modules for Experiments in Stellar Astrophysics (MESA; \citealt{Paxton2015}). We use the test suite case \texttt{1M\char`_pre\char`_ms\char`_to\char`_wd} changing only the mass and turning off RGB and AGB wind mass loss. The CMD location of the primary is fainter than expected from MESA models for a giant with a mass and radius within $1\sigma$ of the asteroseismic measurements. To bring the photometry and asteroseismology into agreement requires using a distance at the highest range of those determined in the literature for M67 (900 pc), and a primary mass and radius of $\sim2.4$ \Msolar~and $\sim8.8$ \Rsolar, more than $2\sigma$ below the asteroseismic values. The BSS and MSTO secondary fall on evolutionary tracks for a $\sim1.4$ and $\sim1.3$ \Msolar~star, respectively. 

While using a distance of 900 pc to M67 creates the best agreement between the asteroseismic mass and the photometry for S1237, \citet{Stello2016b} find an asteroseismic distance to M67 of $816 \pm 11$ pc based on the entire sample of red giants in M67. Using 816 pc results in a photometric mass and radius for the primary more than $3\sigma$ below the asteroseismic values. Depending on the distance used, the asteroseismic results may be $15-30\%$ and $\sim5\%$ higher than the photometric mass and radius, respectively. 

Some recent studies comparing \textit{Kepler} masses and radii from uncorrected asteroseismic scaling relations and RGB eclipsing binaries have found systematic differences between the two measurements (e.g. \citealt{Brogaard2016}, \citealt{Gaulme2016}) similar to the size of the discrepancy we find. We have used corrected scaling relations, which should provide better agreement \citep{Sharma2016}, and \citet{Stello2016b} find that the average RGB mass in M67 from corrected scaling relations is consistent with eclipsing binary results for the cluster. However, the scaling relations have not been well validated by independent measurements (e.g. comparison to eclipsing binaries or cluster isochrones) for HeB stars at masses as large as 2.9 \Msolar, and it is therefore possible that the corrections that work for lower mass stars can not be extrapolated to stars of such high mass. This uncertainty is not reflected in our errors. We encourage more tests of the scaling relations for massive HeB stars to shed light on this possible discrepancy. 

%A first test may be found in the open cluster NGC 6811, which has HeB mass measurements from \text{Kepler} of 2.35 \Msolar  \textbf{reference?} that could be compared to isochrone fitting. 

%This offset indicates that the asteroseismic masses and radii are overestimated by $15-25\%$ and $5-9\%$ respectively (Gaulme et al. 2016, in prep), similar to the discrepancy we find.
Alternatively, S1237 may be following a non-standard evolutionary track due to past mass transfer, a merger, or a stellar collision. However, models of the evolution of mass-transfer products predict they return to close-to-normal evolution soon after mass transfer has ceased \citep{Tian2006}. Models of stellar collision products are bluer and brighter than standard evolutionary tracks due to enhanced mixing enriching the star in helium, although this difference is minor on the giant branch \citep{Sills1997, Sills2009}. Our results indicate the opposite: the star is less luminous and redder than expected given its asteroseismic mass. APOGEE C, N, and Fe abundances do not appear unusual, but a more careful abundance analysis may prove interesting. 

\section{Formation Pathways \label{Formation}}
%Three models have been suggested to explain blue straggler formation: (1) Mass transfer in a binary system, (2) merger of an inner binary in a hierarchical triple due to the Kozai mechanism, and (3) collisions due to dynamical encounters. We explore the viability of each of these three mechanisms to create the S1237 

\subsection{Time Since Formation}
Using MESA we estimate the primary's age to be 450 Myr if it is a 2.9 \Msolar~HeB star. This is likely an overestimate, as it assumes the primary begins with a core that has not burned through any hydrogen. In reality, this star likely formed by adding material to a lower mass MS star to create a BSS. The MS progenitor would already have gone through $\sim$3.5 Gyr of evolution, perhaps resulting in a core substantially depleted of hydrogen. This translates to a shorter lifetime for the star once it becomes a BSS.

If we consider a 2.4 \Msolar~star, $\sim$2 $\sigma$ lower than the asteroseismic value but consistent with the mass implied from its CMD position, the lifetime is $\sim$750 Myr. This longer lifetime would result in a more massive MSTO at the time of formation, and the lower mass makes it more plausible that the star formed from mass transfer or a merger. We discuss this in greater detail in the following sections.

\subsection{Mass Transfer}
The M67 MSTO is 1.3 \Msolar ($1.31 \pm .05$ \Msolar; \citealt{Gokay2013}; \citealt{Stello2016b}), and would have been only  slightly more massive 450 Myr ago. Mass transfer between two such stars would not create a star within the $1\sigma$ uncertainty on the mass of S1237 ($2.97 \pm 0.24$ \Msolar). If we consider a $2.4$ \Msolar~primary the star could have formed from two 1.4~\Msolar~stars (the turnoff mass 750 Myr ago) and a mass transfer origin is worth considering.

The result of mass transfer from an AGB star would be a blue/yellow straggler-WD binary, and thus some mass would be locked up in a WD companion. Given the typical carbon-oxygen WD mass of 0.5-0.6 \Msolar, we would expect a primary with $M \leq 2.3$ \Msolar, even under the unrealistic assumption of totally conservative mass transfer. A lower mass helium WD is not possible because the orbital period of 698 days is too wide to be the result of RGB mass transfer \citep{Rappaport1995}. Additionally, the SED shows no clear evidence of a hot WD companion, though an older ($t\gtrsim 300$ Myr), cool WD would be compatible with the FUV photometry. The system requires a MS companion to explain the SED, so including a WD would require the system to be a triple, e.g. a giant- WD binary with a wide MSTO tertiary. The orbital solution and residuals show no indication of perturbation by such a third body.

Mass transfer leading to a merger between two MS stars could create a $\sim2.6-2.7$~\Msolar~star without a WD remnant. Such mergers may happen when magnetic braking shrinks the orbit of a close binary, but MS mergers may also result from angular momentum loss during a dynamical encounter or Kozai-cycle oscillations (see Sections \ref{Dynamical} and \ref{Kozai}).

\subsection{Dynamical Encounters \label{Dynamical}}

\begin{table*}
\begin{center}
\caption{Encounter Probabilities\label{rates}}
\begin{tabular}{cccccccccccc}
\hline
\hline
Mass (\Msolar) & $\tau_{S1237}$ (Myr) & $\tau_{2+2}$ (Myr) & $\Psi_{2+2}$ & $\tau_{1+3}$ (Myr) & $\Psi_{1+3}$ & $\tau_{3+2}$ (Myr) &$\Psi_{3+2}$ & $\tau_{2+1}$ (Myr) &$\Psi_{2+1}$ & $\tau_{3+3}$ (Myr) &$\Psi_{3+3}$\\
2.4 & 750 & 400 & 0.85& 1300 & 0.44 & 890 & 0.57 & 940 & 0.55 & 3300 & 0.20\\
2.9 & 450 & 400 & 0.68 & 1300 & 0.29 & 890 & 0.40 & -- &-- & 3300 & 0.13\\
\hline
\hline
\end{tabular}
\end{center}
\end{table*}

At least three turnoff mass stars are needed to make a 2.9 \Msolar~primary, and another is required for the secondary. A binary-binary, single-triple, binary-triple, or triple-triple encounter
could provide the four stars needed to create S1237. If we consider a 2.4 \Msolar~primary it is possible to form the primary from just 2 stars,
making a single-binary encounter viable.

\citet{Leigh2011} present a method to estimate BSS production rates from stellar collisions. We use their equations and parameters for M67 to calculate encounter rates and present them in Table~\ref{rates}. We also list
the Poisson probability ($\Psi$) that at least one encounter of each type
would have occurred within the lifetime of S1237, calculated using: 

\begin{center}
\begin{equation}
\Psi_{n+m}= 1-e^\frac{-\tau_{S1237}}{\tau_{n+m}}
\end{equation}
\end{center} Here $\tau_{S1237}$ is the lifetime from Section 5.1, and $\tau_{n+m}$ is the encounter rate.

These rates suggest a few encounters can have occurred within the lifetime of S1237, so \citet{Leigh2011} indicate a collisional origin is possible. We note that post-encounter binaries most commonly have higher
predicted eccentricities than S1237 \citep{Fregeau2004}. Theory predicts S1237 should not have tidally circularized
\citep{Verbunt1995}, so if S1237 was collisionally formed the low eccentricity
is unusual. 

One collision is sufficient to create a $M < 2.7$ \Msolar~primary, but to create a 2.9~\Msolar~primary
requires at least 2 collisions to occur during the encounter. Scattering
experiments show that binary-binary encounters have significant probability
($\sim5-10 \%$) of producing two or more collisions due to the increased
cross-section for interaction after the adiabatic expansion of
the initial collision product \citep{Fregeau2004}.

 Alternatively, S1237 could have undergone multiple collisional
encounters. N-body simulations of M67 have produced multiple `super-BSSs' with masses more than twice the turnoff mass after back-to-back dynamical encounters \citep{Hurley2001}.
While the rates suggest such massive BSSs should be rare, we know
they are produced in open clusters. In addition to S1237, M67
hosts BSSs S977 (a.k.a. F81) and S1082 \citep{Sanders1977}. S977 has a CMD location indicative of
a star more than twice the turnoff mass, and S1082 is a triple BSS system with
a combined dynamically-measured mass of $\sim5.8$~\Msolar~\citep{Sandquist2003}. Multiple collisions
seems to be the most plausible mechanism to create these massive systems.

\subsection{Kozai mechanism}\label{Kozai}
Studies \citep{Perets2009, Naoz2014} have proposed that Kozai-cycle-induced mergers of an inner binary in a hierarchical triple may form BSSs. 

\citet{Naoz2014} run a large set of Monte Carlo models to determine the effect of Kozai cycles on the final orbital configuration of hierarchical triples. They find the triples that lead to inner binary mergers resemble S1237 in several ways. First, they find that mergers most often originate in triples with outer orbital periods ($P_\text{out}$) of $10^3-10^5$ days, consistent with the 698-day period of S1237. Second, they find a peak in the mass ratio ($\frac{m_3}{m_1+m_2}$) between the tertiary and inner binary of 0.4-0.5. Assuming $m_1+m_2= 2.9$ \Msolar~and $m_3= 1.3$ \Msolar, S1237 has a ratio of 0.45. Third, merged systems with $P_\text{out} < 3000$ days have a fairly flat eccentricity distribution from 0.0 to 0.5 \citep{Naoz2014}, consistent with the low orbital eccentricity of S1237 (e= 0.087). 

The vast majority of these systems merge in less than 10 Myr \citep{Naoz2014}. Therefore, if S1237 was a primordial triple we would expect it to have undergone a Kozai induced merger and evolved to a stellar remnant long ago. However, an initially stable primordial triple may be perturbed by a passing star leading to later-in-life Kozai oscillations. Alternatively, a hierarchical triple could have formed dynamically a few hundred Myr ago, and quickly undergone a Kozai induced merger to create S1237. In either scenario, the maximum mass for the merger product would be twice the turnoff mass, or $2.6-2.7$ \Msolar, just outside the $1\sigma$ error on the asteroseismic mass. To produce a 2.9 \Msolar~star would require that one of the stars in the inner binary was already a BSS. 

%We conclude that the orbital properties of S1237 are consistent with those expected for a binary left after a Kozai-induced merger of an inner binary in a hierarchical triple. Due to the short Kozai timescale, this triple would have to have been formed or destabilized in a recent dynamical encounter. 

\section{Summary and Discussion}

We present asteroseismic measurements of a binary YSS in M67 from \textit{Kepler} K2 data. Using $\Delta \nu$-corrected scaling relations, we find the primary star is a 2.9 \Msolar, 9.2 \Rsolar, helium-burning star. This is more than twice the mass of the MSTO. 

SED fitting determines the secondary is near the MSTO, either a normal MS star or a BSS. The SED and stellar models are only marginally consistent with asteroseismology, requiring a primary mass and radius 2-3 $\sigma$ lower than the asteroseismic measurements. This suggests that S1237 is either redder and less luminous than a standard for its mass, or the asteroseismic mass and radius are too large by at least $\sim20$ and $\sim5\%$, respectively. We encourage testing the asteroseismic scaling relations for massive helium burning stars for possible discrepancies. 

Observed X-ray emission from this star has not been explained. Given the masses of the primary and secondary $M_1$= 2.9 \Msolar, $M_2$=1.3 \Msolar), the mass function indicates the orbit is inclined at $\sim20^{\circ}$. If the spin and orbital axes are aligned, the giant primary may be
rapidly rotating without observable line broadening or seismic frequency splitting \citep{Beck2012}, thereby explaining the
X-rays. Alternatively the X-ray emission may arise from a rapidly rotating
secondary. Many X-ray sources in M67 with similar X-ray luminosities are
rapid rotators located near the MSTO (see
Fig.~\ref{CMD}). Most are tidally synchronized close binaries, which is not the case for S1237, but rapid rotation is also expected after a recent
stellar collision or mass transfer event.

We review possible formation models and suggest that a binary encounter
leading to one or more collisions or mergers is a possible formation pathway. Collisions may have taken place during the dynamical encounter itself, or perhaps a Kozai-induced merger of an inner binary in a dynamically formed triple occurred shortly afterward. 

This is the first asteroseismic mass measurement for a YSS and the second YSS for which a formation pathway has been suggested. S1040, another YSS in M67, has a helium WD companion, suggesting it formed through mass transfer \citep{Landsman1997}. These two examples indicate that YSSs are likely evolved BSSs, and that like the BSSs there are multiple mechanisms to create them.
% \textbf{Referee's comments about the last line "the statement is strong and misleading ('This is the first YSS with a directly measured mass,') because we still not sure the implication of using the scaling relation for a HeB star with such a mass. And as the work shows, it is incompatible with other methods (CMD)." Any suggestions for how to end}

\acknowledgements
This paper includes data collected by the Kepler mission. Funding for the Kepler mission is provided by the NASA Science Mission directorate.

%This publication makes use of data products from the Two Micron All Sky Survey, which is a joint project of the University of Massachusetts and the Infrared Processing and Analysis Center/California Institute of Technology, funded by the National Aeronautics and Space Administration and the National Science Foundation.

%This publication makes use of data products from the Wide-field Infrared Survey Explorer, which is a joint project of the University of California, Los Angeles, and the Jet Propulsion Laboratory/California Institute of Technology, funded by the National Aeronautics and Space Administration.

%This project uses observations made with the NASA Galaxy Evolution Explorer. GALEX is operated for NASA by the California Institute of Technology under NASA contract NAS5-98034.

E.L. and R.M. are supported by NASA Grant NNX15AW69G. E.L. is also funded by the Wisconsin Space Grant Consortium. A.V. is supported by the NSF Graduate Research Fellowship, Grant No. DGE 1144152. 
\bibliographystyle{mn2e}
%\bibliography{ngc6791.ssg}

\end{document}